\documentclass[aps,prd,10pt,         
               preprintnumbers,numbers,sort&compress,nofootinbib,
                            showpacs,
               colorlinks,
               linkcolor=blue,
               citecolor=blue]{revtex4-1}
\usepackage{amsmath}
\usepackage{amsmath}
\usepackage{psfrag}
\usepackage{epsfig}
\usepackage{graphicx}
\usepackage{epstopdf}
\usepackage{graphicx,amsmath,amssymb,bm}
\usepackage{psfrag}
\usepackage{feynmp}
\usepackage{hyperref}
\usepackage{enumitem}

\newcommand{\exclude}[1]{}

\begin{document}
\newcommand{\beq}{\begin{equation}}
\newcommand{\eeq}{\end{equation}}
\def\Journal#1#2#3#4{{#1} {\bf #2}, #3 (#4)}

\newcommand{\A}{\alpha}
\newcommand{\B}{\beta}
\newcommand{\T}{\theta}
\newcommand{\Ep}{\epsilon}
 \newcommand{\fr}{\frac}
\newcommand{\be}{\begin{eqnarray}}
\newcommand{\ee}{\end{eqnarray}}
\newcommand{\G}{\gamma}
\newcommand{\D}{\delta}
\renewcommand{\P}{\phi}
\newcommand{\intl}{\int\limits_{1}^{\infty}}
\renewcommand {\L}{\lambda}
\newcommand{\pt}{\partial}

\newcommand{\bq}{\bar q_A}
\newcommand{\tq}{\tilde q_A}
\newcommand{\btq}{\bar{\tilde q}^A}
\newcommand{\fa}{\varphi^A}
\newcommand{\bfa}{\bar \varphi_A}

\newcommand{\none}{{\cal N}=1}                            
\newcommand{\ntwo}{{\cal N}=2}                            

\def\st{\scriptstyle}
\def\sst{\scriptscriptstyle}
\def\mco{\multicolumn}
\def\epp{\epsilon^{\prime}}
\def\vep{\varepsilon}
\def\ra{\rightarrow}
\def\al{\alpha}
\def\ab{\bar{\alpha}}
 \def\dd{ \,\mathrm{d} }
\def\d{\partial}
\def\+{\dagger}
\def\la{\langle}
\def\ra{\rangle}
\def\<{\langle}
\def\>{\rangle}
\def\gmmu{\gamma _{\mu}}
\def\gmnu{\gamma_{\nu}}
\def\atop{\frac{ \alpha_{s}}{8 \pi} G_{\mu \nu}^{a}
 \tilde{G}^{\mu \nu a} }
\def\gmf{\gamma _{5}}
\newcommand{\Lqcd}{\Lambda_{\mathrm{QCD}}}
\newcommand{\Lbar}{\Lambda_{\overline{\mathrm{QCD}}}}
\newcommand{\qcd}{{\overline{\mathrm{QCD}}}}
\newcommand{\mq}{m_{\overline{q}}}
\newcommand{\Tr}{\mathrm{Tr}}
\newcommand{\cph}{\varphi}
\newcommand{\cth}{\vartheta}

\title{  Contact Term,  its  Holographic Description  in QCD and Dark Energy}

 \author{Ariel R. Zhitnitsky}

\affiliation{Department of Physics \& Astronomy, University of
  British Columbia,  Vancouver,  BC  V6T1Z1,Canada}
 \date{\today}

\begin {abstract}
In this work we 
study the  well known contact term, which is the key element in resolving the so-called $U(1)_A$ problem in QCD. 
We study this term using the dual   Holographic Description.   We argue that in the dual picture  
the contact term is saturated by the D2 branes which can be interpreted as the tunnelling events in Minkowski space-time.
We quote  a number of direct lattice results supporting this identification. We also argue that the contact term 
receives a Casimir -like correction $\sim (\Lqcd R)^{-1}$ rather than naively expected $\exp(-\Lqcd R)$ when  the Minkowski space-time ${\cal R}_{3,1}$ is replaced by a large but finite manifold with a size $R$. Such a behaviour is  consistent  with  other QFT-based  computations
when power like corrections are due to nontrivial properties of topological sectors of the theory.   In holographic description such a behaviour is due to massless Ramond-Ramond (RR) field  living in the bulk of multidimensional space when  power like corrections  is a  natural outcome of massless  RR field. In many respects the phenomenon  is similar to the Aharonov -Casher effect when the ``modular electric field" can penetrate into a superconductor where the electric field is exponentially screened.   The role of ``modular operator" from Aharonov -Casher effect  is played by 
large gauge transformation operator    $\cal{T}$   in 4d QCD, resulting the  transparency of the system to   topologically nontrivial  pure gauge configurations.
 We discuss some    profound consequences of our findings.    In particular, we speculate that a slow  variation of the contact term in expanding universe   might be the main source of the observed   Dark Energy. 

\end{abstract}

\maketitle
\section{Introduction}
The holographic picture of QCD is known to provide some deep insights into the strongly coupled dynamics.
The most well known example is the dual model of gluodynamics at zero and nonzero temperatures  \cite{wittenterm}, and its generalization \cite{ss}  when light fermions are included into the system.

The goal of thsi work is to study the well known contact term which is the key element in  in resolving the so-called $U(1)_A$ problem in QCD \cite{witten,ven, vendiv},
 see also~\cite{Rosenzweig:1979ay,Nath:1979ik,Kawarabayashi:1980dp}. We want to understand some deep properties of the contact term from the dual perspective.

 The basic tools in this study will be the D branes which are the crucial elements of the dual description. 
 It is well known that D0 brane extended along $x_4$ can be  identified with  conventional  instanton
\cite{Bergman:2006xn}. At the same, the D2 brane wrapped around both periodic
coordinates was identified with  magnetic string \cite{Gorsky:2007bi}, while D2 brane wrapped around 
$x_4$ was identified with D2 domain wall~\cite{Yee}.  These two branes will play an important role in our studies. 
For completeness, we should also mention the configuration of 
the D6 brane wrapped
around the compact $S^4$ part of the dual geometry  was interpreted as the domain wall separating two vacua
in pure gluodynamics \cite{wittenflux}. The corresponding generalization with quarks was given  in  
 \cite{Hong:2008nh,Lin:2008vv}.     A large number of  other D defects from this Zoo have been discussed in \cite{Gorsky:2009me}.

 The crucial element which helps to make the identifications is the study of the confinement- deconfinement phase transition which is interpreted in the dual picture as  the Hawking-Page phase transition \cite{wittenterm}.  
In the dual picture 
the wrapping around
$x_4$ is stable above the phase transition $T>T_c$  as a result of cylinder geometry, while it is    unstable below the critical temperature
$T<T_c$ as a result of cigar type geometry, see definitions and details below. The wrapping around $\tau$ does the opposite: the corresponding D brane  is stable at small
temperatures $T<T_c$ and unstable at high temperatures, $T>T_c$.
 Since the contact term  is sensitive to the $\theta$-parameter we shall also
discuss the $\theta$ dependence  from the dual perspective.

  To formulate the problem we start from the conventional definition of the topological susceptibility $\chi$
 which plays a crucial role in resolution of the $U(1)$ problem\cite{witten,ven,vendiv,Rosenzweig:1979ay,Nath:1979ik,Kawarabayashi:1980dp} 
\be
\label{chi}
 \chi (\theta=0)= \left. \frac{\partial^2E_{\mathrm{vac}}(\theta)}{\partial \theta^2} \right|_{\theta=0}=  \lim_{k\rightarrow 0} \int \!\dd^4x e^{ikx} \la T\{q(x), q(0)\}\ra ,
 \ee
where  $\theta$ is the  $\theta$ parameter which enters the  Lagrangian along with  topological density operator $q (x)$, see precise definitions below. 

It is important  that the topological susceptibility $\chi$  does not vanish in spite of the fact that $q= \partial_{\mu}K^{\mu}$ is total divergence.  Furthermore, any physical asymptotic states gives a negative contribution\footnote{We use the Euclidean notations  where  path integral computations are normally performed.}
 to this 
diagonal correlation function
\be	\label{G}
  \chi_{\rm dispersive} \sim  \lim_{k\rightarrow 0} \int \!\dd^4x e^{ikx} \la T\{q(x), q(0)\}\ra  \sim  
    ~\lim_{k\rightarrow 0}   \frac{\la  0 |q|G\ra \la G| q| 0\ra }{-k^2-m_G^2}\simeq -\frac{|c_G|^2}{m_G^2} \leq 0.
\ee
 where   $m_G$ is the mass of asymptotic state,  $k\rightarrow 0$  is  its momentum, and $\la 0| q| G\ra= c_G$ is its coupling to topological density operator $q (x)$.
 At the same time the resolution of the $U(1)_A$ problem requires a positive sign for the topological susceptibility (\ref{top1}), see the original reference~\cite{vendiv} for a thorough discussion, 
\be	\label{top1}
  \chi_{\rm non-dispersive}= \lim_{k\rightarrow 0} \int \!\dd^4x e^{ikx} \la T\{q(x), q(0)\}\ra > 0 \, .
\ee

Therefore, there must be a contact contribution to $\chi$, which is not related to any propagating  physical degrees of freedom,  and it must have the ``wrong" sign. The ``wrong sign" in this paper implies a sign 
  which is opposite to any contributions related to the  physical propagating degrees of freedom.   In the framework \cite{witten} the contact term with the ``wrong sign" has been simply postulated, while in refs.\cite{ven,vendiv} the Veneziano ghost had been introduced to saturate the required property (\ref{top1}).   These two descriptions are equivalent  as they describe the same physics.   
  
It interesting to note that the ``wrong" sign in topological susceptibility (\ref{top1}) is not the only manifestation of this ``weird" unphysical  degree of freedom. In particular, one can argue that this term also contributes with a wrong sign into the entropy. While the entropy itself is a positively defined entity, the corresponding gauge invariant contribution  
related to the  topological susceptibility contribute with the negative sign ~\cite{Zhitnitsky:2011tr,Donnelly:2012st}.

 The goal of this paper is to investigate this ``weird term"  using the holographic description. 
 We also want to study the behaviour of the system when size of then system is large  but not infinitely 
large. The last property, as we argue below,  might be relevant for cosmological applications, see section \ref{consequences}. We also want to study the correlation between drastic changes in $\theta$ behaviour  when the phase transition is crossed at $T_c$. This point has been emphasized in \cite{Bergman:2006xn, Gorsky:2007bi, Parnachev:2008fy,Gorsky:2009me} from holographic perspective as well as  from quantum field theory (QFT) viewpoint. 
This correlation is also supported by direct lattice studies, see review  \cite{Vicari:2008jw}
and references on the original papers therein.

The paper is organized as follows. 
In section \ref{QCD} we overview  the nature of the contact term from QFT viewpoint.  
In Section \ref{model} we describe
our model based on the $N_c$ D4 branes with one compact world-volume coordinate.
We also overview some  D- defects relevant for our analysis. We treat D defects  in the probe approximation
when they do not deform the dual geometry.  
In Section \ref{holography}   we formulate our proposal on contact term from holographic perspective. 
We present a number of arguments supporting our construction, including the comparison with the direct lattice measurements.
We also argue in section \ref{corrections} that the contact term 
receives a Casimir -like correction $\sim (\Lqcd R)^{-p}$ rather than naively expected $\exp(-\Lqcd R)$ when  the Minkowski space-time ${\cal R}_{3,1}$ is replaced by a large but finite manifold with a size $R$.   In holographic description such a behaviour is due to massless Ramond-Ramond (RR) field  living in the bulk of multidimensional space when  power like corrections  is a  natural outcome of massless  RR field. In many respects the phenomenon  is similar to the Aharonov -Casher effect when the ``modular electric field" can penetrate into a superconductor where the electric field is exponentially screened.    In Section \ref{consequences} we outline some profound consequences of our findings. In particular, we speculate that a slow  variation of the contact term 
in expanding universe   might be the main source of the so called  Dark Energy (DE).

\section{Contact term in QCD from QFT viewpoint. }\label{QCD}
In this section we want to provide some intuition on the nature of the contact term using some model computations. 
The simplest possible example is two dimensional $QED_2$ in  the  Kogut-Susskind formulation~\cite{KS} where all essential elements  related to the contact term can be explicitly worked out as discussed in refs.~\cite{Zhitnitsky:2010ji, Zhitnitsky:2011tr}. However, for  purposes of the present work we present  the relevant  exact results for four dimensional ``deformed QCD" formulated in \cite{Yaffe:2008} where all computations can be explicitly performed as the model is in a weak coupling regime, see  next subsection \ref{deformed}. We reformulate the same results for  strongly coupled 4d QCD using effective description in terms of the Veneziano ghost as originally developed in~\cite{ven,vendiv}, see section \ref{ghost}. 

\subsection{ Contact term in weakly coupled ``deformed QCD"}\label{deformed}

In the deformed theory an extra term is put into the Lagrangian in order to prevent the center symmetry breaking\cite{Yaffe:2008}. It can be arranged in such a way 
the constructed theory    remains confined at high temperature in a weak coupling regime, 
such that  there is no  order parameter to differentiate the low temperature   confined regime from the high temperature   confined regime. Therefore, one can use this weakly coupled gauge theory to test some deep theoretical ideas
 about confined phase.   
  
The topological susceptibility in this model can be explicitly computed and is given by \cite{Thomas:2011ee}
\begin{equation} \label{chi_QCD}
	\chi_{QCD} = \int d^4 x \< q(\bold{x}) q(\bold{0}) \>  =\frac{\zeta}{N_cL} \int d^3 x \left[ \delta(\bold{x})
		-m_{\eta'}^2 \frac{e^{-m_{\eta'}r}}{4\pi r}  \right],
\end{equation}
where the monopole fugacity $\zeta$  can be explicitly computed in this model \cite{Yaffe:2008}.  The monopoles are pseudoparticles in this model. Therefore,  the monopole fugacity $\zeta$  should be understood as number  of  tunnelling events per unit time per unit volume. 
There are two terms in  formula (\ref{chi_QCD}). The first term represents the non-dispersive contribution not related to any physical propagating degrees of freedom, while the second terms represents the conventional $\eta'$ contribution. 
As one can see the first term contributes with sign plus while the second term contributes with sign minus   in accordance with general equation (\ref{G}). Another important  lesson from this equation: the Ward Identity (WI) expressed as  $\chi_{QCD} (m_q=0)=0$ is automatically satisfied as a result of cancellation between the two terms.

It is important to note  that the number of tunnelling events per unit time per unit volume (\ref{chi_QCD}) in pure gauge  theory in this model (with no  quarks) is related   with the absolute value of the energy density of the system.  Indeed,  
\be
\label{vac}
E_{\mathrm{YM}}(\theta)=- \frac{N_c\zeta}{L}\cos\left(\frac{\theta}{N_c}\right), ~~~\chi_{YM} (\theta=0)=  \left. \frac{\partial^2E_{\mathrm{YM}}(\theta)}{\partial \theta^2} \right|_{\theta=0}=\frac{\zeta}{N_cL}, 
\ee
where we keep only the lowest branch $l=0$ in expression for  $\cos\left(\frac{\theta +2\pi l}{N_c}\right)$ to simplify formula (\ref{vac}),
see detail discussions with complete set of references on this matter in  \cite{Thomas:2011ee}. In different words, the contact term in pure gauge theory $ \chi_{YM}  =\frac{\zeta}{N_cL}$ can be interpreted in terms of number tunnelling events between   different topological sectors in the system. Therefore, there is no surprise that it has ``wrong sign" as the relevant physics 
can not be described in terms of propagating physical degrees of freedom, but rather, is described in terms of the tunnelling events between  different (but physically equivalent) topological sectors in the system.

  \subsection{Contact term in terms of the Veneziano ghost in strongly coupled QCD} \label{ghost}
  Our discussions thus far were limited to the model where all computations are justified
as a result of  the weak coupling regime enforced by deformation \cite{Yaffe:2008}.
  Now we want to explain the same physics, the same contact term with ``wrong sign"  but in different way. 
  We want to formulate this  physics in terms of the Veneziano ghost which effectively describes the dynamics of the topological sectors in strongly coupled regime. Such an alternative description in terms of the Veneziano ghost  ~\cite{ven,vendiv}  will play an important role in our identification of the relevant configurations in next section when we will discuss the contact term using  the dual holographic description.

The topological susceptibility    in the chiral limit  can be  easily computed in the model  ~\cite{ven,vendiv} using the Veneziano ghost 
and it is given by ~\cite{dyn,Zhitnitsky:2010zx}:  
   \be
\label{top2}
\chi_{QCD} &\equiv& \int \!\dd^4x \la T\{q(x), q(0)\}\ra  
  = \frac{f_{\eta'}^2 m_{\eta'}^2}{4} \cdot \int d^4x\left[ \delta^4 (x)- m_{\eta'}^2 D^c (m_{\eta'}x)\right].
\ee
 The Green's function   $ D^c (m_{\eta'}x)$ in this expression  describes  free (in the chiral limit) massive $\eta'$ field  and satisfies  standard normalization $\int d^4x m_{\eta'}^2 D^c (m_{\eta'}x)=1$. 

The structure of this expression (\ref{top2}) is   identical to our formula   (\ref{chi_QCD}) computed in weakly coupled ``deformed QCD".
 In particular, the term proportional $ -D^c (m_{\eta'}x)$ with negative sign  in eq. (\ref{top2})   is  resulted from the lightest  physical $\eta'$ state  of   mass $m_{\eta'}$. At the same time, the contribution with ``wrong sign" expressed  by 
 $\delta^4(x)$ in eq. (\ref{top2}) is   nothing else but the contact term  
 \be
\label{YM}
\chi_{YM} &\equiv& \int \!\dd^4x \la T\{q(x), q(0)\}\ra_{YM}  =\frac{f_{\eta'}^2 m_{\eta'}^2}{4} \cdot \int d^4x\left[ \delta^4 (x)\right].
\ee
 In  the model  ~\cite{ven,vendiv}  this term is saturated by the Veneziano ghost  contribution.   The required ``wrong sign" for this contribution is due to the negative sign  
 for the kinetic term for the Veneziano ghost.
In our weakly coupled  ``deformed QCD" when all tunnelling processes can be explicitly computed this contact   term in eq. (\ref{chi_QCD})  is  expressed in terms of the monopole's fugacity $N_c\zeta/L$, and describes tunnelling  transition between the topological sectors of the theory. It is natural to expect that the same interpretation should be also applied to dimensional     factor  $\frac{f_{\eta'}^2 m_{\eta'}^2}{4}$ which enters (\ref{YM}). In different words, this factor  should be also interpreted in terms of   tunnelling   of some  ``objects". However, the nature of these objects can not be easily identified in contrast with weakly coupled ``deformed QCD" where the relevant pseudo-particles are obviously the monopoles saturating the contact term.  We shall see however  in next section that such an interpretation  still can be 
 advocated  even   in strongly coupled gauge theory. However,  our  arguments will be  based on dynamics of  ``dual objects" rather than on dynamics  of the original gauge fields.

 One should say that the main  features represented by eqs. (\ref{top2}), (\ref{YM}) such as singular behaviour of the contact term with positive sign, and a smooth behaviour of conventional dispersive contribution are explicitly seen in lattice simulations ~\cite{Horvath:2003yj,Horvath:2005rv,Horvath:2005cv,Alexandru:2005bn}. A similar structure has been also 
observed in QCD by  different groups~\cite{Ilgenfritz:2007xu,Ilgenfritz:2008ia,Bruckmann:2011ve,Kovalenko:2004xm,Buividovich:2011cv} and also in two dimensional $CP^{N-1}$ model~\cite{Ahmad:2005dr}. Most important  part for us from these numerical studies is that  the singular behaviour of the contact term is not an artifact of any approximation, but an inherent feature of underlying gauge theory. 
  Furthermore, there are  no any physical scale factors (such as $\Lqcd$)  which would determine  the  singular behaviour of this term.      In different words, the Veneziano ghost does model 
 the crucial  property of the  topological susceptibility related to summation over topological sectors  in gauge theories. This feature  can not be accommodated by any physical asymptotic states as it is related to  non-dispersive contribution  with ``wrong sign".  
 
     This concludes our discussions of the contact term within QFT approach.  The corresponding  computations give us a hint
on the nature of the contact term, and quantum configurations which saturate it. These insights  will play an important role in our analysis   of the contact  term 
in holographic description, where we attempt to  identify  the relevant   configurations which saturate $\chi_{YM}$ in the dual picture.

     \section{Description of the model}\label{model}
     We start with the Witten's background \cite{wittenterm} for $SU(N_c)$ Yang-Mills theory in large $N_c$ limit, which is the gravity dual to the theory 
      on  $N_c$ D4 branes wrapped around a compact dimension $S^1$.
     In the supergravity  approximation the geometry looks as $M_{10}=R_{3,1}\times D \times S^4$ and
the corresponding metric in notations \cite{Bergman:2006xn} reads as
\be 
\label{metric}
ds^2&=&(\frac{u}{R_0})^{3/2}(-dt^2 + \delta_{ij}dx^i dx^j +f(u)dx_4^2)+
(\frac{u}{R_0})^{-3/2}(\frac{du^2}{f(u)} +u^2 d\Omega_{4}^2)\nonumber \\
e^{\Phi}&=&(\frac{u}{R_0})^4, 
\qquad
f(u)=1- (\frac{u_{\Lambda}}{u})^3, \qquad R_0=(\pi g_s N_c)^{1/3}, \qquad R=\frac{4\pi}{3}(\frac{R_0^3}{u_{\Lambda}})^{1/2}. 
\ee
The coupling constant of Yang-Mills theory and the radius
of the compact dimension $R$ are related as follows
\beq
g_{YM}^2=\frac{8\pi ^2 g_s l_s}{R},~~~~~~  \lambda\equiv\frac{g_{YM}^2 N_C}{2\pi}. 
\eeq
In formula (\ref{metric}) the confined phase corresponds to the geometry
of  a cigar  in the ($u,x_4$) coordinates with the tip at $u=u_{\Lambda}$.
At the same time, in de-confined phase the ($u, x_4$) coordinates  exhibit the cylinder geometry.
The situation is reversed when one considers  ($\tau,u$) plane instead
of ($x_4,u$) plane,  where
$\tau$ is the Wick-rotated time coordinate $\tau=it$,
$\tau\propto \tau + \beta$.  
It has been argued in  \cite{wittenterm} that
the phase transition occurs exactly  when one geometry replaces another. 
Important  feature of this transition is as follows.
The
wrapping around the internal $x_4$ circle is topologically stable,
while the wrapping around the  Euclidean time coordinate $\tau$ 
is unstable above the phase transition point $T>T_c$. The opposite patter occurs of the two wrappings
below the phase transition at $T<T_c$. Namely, the
wrapping around the internal $x_4$ circle is topologically unstable, while the wrapping around the  Euclidean time coordinate $\tau$ becomes stable at $T<T_c$. 

We want to discuss the topological features of the theory related to the  $\theta$-dependence.
The corresponding bulk field that couples to the topological density operator  
is the  Ramond-Ramond (RR) field. The precise relation between $\theta$ parameter and RR field is known, 
\begin{equation}
\label{RR}
\theta =\frac{1}{l_s} \int dx_4\ C_1
\end{equation}
where  $C_1$ is the RR  1-form  field. 
Important consequence of the relation (\ref{RR}) is that 
 the $\theta$ dependence of the worldvolume
theories on the D branes  correlates with the wrapping around $x_4$ coordinate.
In different words, D brane configurations might be relevant for study the $\theta$ dependence if the corresponding construction includes the wrapping around $x_4$ coordinate. Otherwise, the D brane configurations can not be sensitive to the $\theta$ parameter. 
 
We conclude this section with short description of D branes which will play an important  role in our discussions which follow, 
see  \cite{Gorsky:2009me} with description  of other members of this Zoo:

 {\it \underline{D0 instantons}.}\\
The D0 brane extended along $x_4$ was identified as the YM instanton
\cite{Bergman:2006xn}. At $T>T_c$ this wrapping is topologically stable, while at $T<T_c$ 
the D0 brane tends to shrink to the tip
where its tension vanishes. Nevertheless, it has been argued in \cite{Parnachev:2008fy,Gorsky:2009me} 
that the D0 branes  might be  still important configurations at $T<T_c$. 

{\it \underline{D2 string}.}\\
The magnetic string
is the probe D2 brane wrapped
around $S_1$ parameterized by $x_4$ and its tension
is therefore proportional to the effective radius $R(u)$
\cite{Gorsky:2007bi}.
At small temperatures   this wrapping
is topologically unstable and the D2 brane tends to shrink to the tip
where its tension vanishes. 
 The magnetic string carries  nontrivial 4d topological charge.
 
 {\it \underline{D2  domain wall
}.}\\
The D2 brane wrapped around 
$x_4$ was identified as D2 domain wall~\cite{Yee}. At small temperatures   this wrapping
is topologically unstable and the D2 branes tend to shrink to the tip
where their tension vanish. At  $T>T_c$ the D2 branes are  topologically stable objects and have  been  identified 
with $Z_N$ domain walls which are present in the system in high temperature phase.

 {\it \underline{D6  domain wall
}.}\\
If D6 brane   wraps around $\tau $ it behaves as the domain wall
which is a source of the corresponding RR-form. Such a configuration has been interpreted
in ref. \cite{wittenflux} as the domain wall which separates
different metastable vacua  known to exist in gluodynamics at large $N_c$.
Its worldvolume theory on the domain wall   has no
$\theta$ dependence as D6 brane   wraps around $\tau $ rather than around $x_4$ which was the case for D0, D2 branes mentioned above.

\section{Holographic description  of the contact term}\label{holography}
The computation of the contact term in holographic picture is known~\cite{wittenflux,Hashimoto:1998if}, see also review paper \cite{DiVecchia:1999yr}.
The contact  term has a positive sign and has expected $N_c^0$ behaviour in accordance with QFT expression (\ref{YM}).
The final formula in  holographic computations is expressed as a surface integral in 5-th dimension   and in notations of section \ref{model}  is given by ~\cite{wittenflux,Hashimoto:1998if,DiVecchia:1999yr}:
\be
\label{du}
\chi_{YM}\sim \int_{u_{\Lambda}}^{\infty}d u \partial_u \left[ \sqrt{g} g^{x_4 x_4}g^{uu}h(u) \partial_u h(u)\right]\sim M^4 \lambda^3 N_c^0, ~~ h(u)\sim f(u)= \left[1- \left(\frac{u_{\Lambda}}{u}\right)^3 \right],
\ee
where $M=[\oint dx_4]^{-1}$ is the only dimensional parameter of the problem, to be identified with $\Lambda_{QCD}$. Function 
  $h(u)$ in (\ref{du}) is $4-th $ component of the  R-R   field $C_1$ which satisfies classical equation of motion
\be
\label{laplace}
  \partial_{\mu} \left[ \sqrt{g} g^{x_4 x_4}g^{\mu\nu}  \partial_{\nu} h(u)\right]=0. 
\ee
 The  solution of this equation is given by expression (\ref{du}), see \cite{wittenflux,Hashimoto:1998if,DiVecchia:1999yr} for the technical details.

We have nothing new  to add to those computations. The goal of this section is in fact quite different: we want to understand the physics 
which is hidden behind these computations. In different words, we want to 
understand the relation between the holographic formula (\ref{du}) and the QFT based computations (\ref{top2}, \ref{YM}).  As we mentioned previously, the QFT based computations (\ref{top2}, \ref{YM})   are supported by the lattice studies ~\cite{Horvath:2005cv,Ilgenfritz:2007xu, Ilgenfritz:2008ia,Bruckmann:2011ve}, including the singular behaviour of the contact term with vanishing size in the continuum limit.  This singular behaviour,  as it is known, leads to  a finite integral contribution  to $\chi_{YM}$ with the ``wrong sign"  in agreement with  eqs. (\ref{top2}, \ref{YM}). The understanding of this singularity  in the dual description will be the crucial element in our analysis  on a possible variations of the contact  term  when 
the background slightly varies, and when formulae similar to (\ref{du}, \ref {laplace}) are not presently available. Our findings  regarding these possible  variations  with slight variation of a background will play a key role in the  application considered in section \ref{consequences}
where we interpret a tiny  deviation of the $\theta -$ dependent portion of the vacuum  energy  (coded by $\chi_{YM}$)
in expanding universe as a main source of the observed dark energy.    
 \subsection{Emergence of the contact  term as a tunnelling process.  Percolation of D2 branes. } \label{lattice} 
 The presence of non-dispersive contact term (\ref{YM})  in topological susceptibility in pure gluodynamics 
 obviously implies that there must be unphysical pole in gauge variant correlation function  
 \be
\label{top3}
\chi_{YM}^{\mu\nu}(k)\sim \lim_{k\rightarrow 0} \int \!\dd^4x e^{ikx}\la T K^{\mu}(x) , K^{\nu}(0) \ra  \sim \frac{k^{\mu}k^{\nu}}{k^4}, ~~~ 
\chi_{YM}\sim k_{\mu}k_{\nu}\cdot \chi_{YM}^{\mu\nu}
\ee
While the correlation function $\chi_{YM}^{\mu\nu}(k)$ is gauge variant itself, the merely presence of the pole, its position, its residue with ``wrong sign" are physically  observable parameters as corresponding correlation function (\ref{chi}), (\ref{YM}) is  gauge invariant object. 
What is the nature of this pole? What it could be? How it can be interpreted from the holographic perspective? Normally, a  pole at zero mass corresponds 
to a  massless gauge boson. Or  it might be  a result of spontaneous symmetry breaking effect. What is a symmetry which could be responsible for behaviour  (\ref{top3})?  Furthermore, as explained  in section \ref{QCD} this pole must have a residue with a ``wrong sign"   such that it can not be identified with any {\it physical} propagating  massless degree of freedom. In QFT  description this pole corresponds to the  contact term  which emerges due  to the presence  of  different topological sectors  
in the system as argued in  section \ref{QCD}. In particular, in weakly coupled ``deformed QCD" the contact term is saturated by monopoles which describe the tunnelling between those topological sectors as discussed in \ref{deformed}.    In strongly coupled regime analytical computations based on underlying gauge fields are not reliable. However,  one can argue    that  the contact term is saturated by the Veneziano ghost as reviewed  in section \ref{ghost}. In this framework the Veneziano ghost  should be treated as an auxiliary unphysical degree of freedom which 
does not belong to the physical Hilbert space.  Nevertheless, it  saturates 
topological susceptibility (\ref{top2}) and describes the dynamics of these degenerate topological sectors of the theory.
     
     The QFT- based insights  presented  above  suggest  that the origin for the contact term in holographic description should  be also interpreted as  a result of tunnelling events rather than presence of any 
     massless propagating degrees of freedom. The tunnelling events in path integral  approach in four dimensional Euclidean space should be manifested   as a presence of some ``objects" which live in four dimensions. We argue below that we can identify these objects as D2 branes wrapped around $S_1$ parametrized by $x_4$.
As we mentioned   in section \ref{model} the tension for the D2 brane     vanishes as a result of cigar geometry such that the D2 brane configuration shrinks to the tip.
Therefore, an arbitrary large number of D2 branes could be produced, resulting in their condensation.  The term ``condensation" we use here is actually a jargon,
as D2 branes are 4d time- dependent objects rather than 3d static objects when term ``condensation" is normally used. The ``tunnelling" or ``4d -percolation" probably  are more appropriate   terms to describe the relevant physics in this case.  However, we keep the term ``condensate" as it is an appropriate term when the system is viewed from 5d perspective,
which is precisely what holographic description is about.  We shall present below a number of arguments supporting this identification.  

To begin with  our arguments, we  remind  the reader the Bloch's theorem in condensed matter physics which states the following. Consider a  system (such as a perfect lattice) which is invariant under  displacement $x\rightarrow x+a$  symmetry. The system in such circumstances can be described by the bands of allowed and forbidden 
energies classified by a quasi-momentum. Furthermore,  a quasi-electron in allowed band is characterized by ungapped dispersion relation $\epsilon (k)=\frac{k^2}{2m^*}$ such that the corresponding correlation function has a pole at $\left[\epsilon (k)-\frac{k^2}{2m^*}\right]=0$ while all information about tunnelling  
properties are coded by parameter $m^*$ and by residue of  the Green's function. The presence of such pole  implies that the corresponding quasi-particle can freely propagate to arbitrary large distances without any disturbance 
from the strongly interacting  lattice  as long as the symmetry $x\rightarrow x+a$ is respected and no any lattice's defects are included into the consideration. It should be contrasted with a case when  an electron is placed into a system with a conventional potential barrier, in which case the correlation function will  be characterized by a gapped dispersion relation with a pole at $[\epsilon (k)=\omega_0+\frac{k^2}{2m^*}]$ such that a particle can propagate only for a short time $\sim \omega_0^{-1}$.

  Our system with winding states $| n\ra$, quasi-momentum $\theta$   and  large gauge transformation operator  ${\cal T}$
   with property  ${\cal T} | n\ra= | n+1\ra$ is analogous to the Bloch's displacement symmetry  $x\rightarrow x+a$ with the ``only" difference that the states $| n\ra$ are not physically distinct states, and there is no any  real ``physical" degeneracy in the system.  The physical vacuum state in QCD  is {\it unique}
     and constructed as a superposition of $| n\ra$ states. Nevertheless, the presence of the pole in eq. (\ref{top3}) can be interpreted as a result of tunnelling      between  topological sectors  $| n\ra$, similar to the tunnelling events in the Bloch's system. It also explains the ``wrong" sign in residues of the correlation function (\ref{top3}) as we describe the tunnelling of the Euclidean D2 objects  between winding states $| n\ra$ rather than the tunnelling of conventional physical degree of freedom between distinct vacuum states  in condensed matter physics in the Bloch's case, when the residue in the Green's function (coefficient in front of the pole) is precisely the probability of the tunnelling. It further supports our interpretation  of the coefficient $ {f_{\eta'}^2 m_{\eta'}^2}/{4}$ in eq. (\ref{YM}) in strongly coupled QCD as a probability of tunnelling of ``some objects" which we identify now with D2 branes. 
     
     Now we can answer the question formulated in the beginning of this section: what is a symmetry which could be responsible for behaviour  (\ref{top3})? It is the invariance under  large gauge transformations defined by the operator  ${\cal T}$ such that $[{\cal T}, H]=0$, see Appendix \ref{tau} for a technical  details. 
     We come back to this  interpretation   at the end of this section when we discuss analogy with  Aharonov -Casher effect
     and similarity between  the large gauge transformation operator  ${\cal T}$ and their ``modular operator"  commuting  with the hamiltonian.

   The tunnel-based 
   interpretation unambiguously implies that the corresponding D-defect which could be  potentially responsible 
   for tunnelling 
   must be constructed with wrapping  around $S_1$ parametrized by $x_4$ rather than by $\tau$ coordinate. In this case  the corresponding  D defect   becomes the $\theta$- dependent object 
   (and therefore becomes sensitive  to the transitions $  | n\ra \rightarrow  | n+1\ra$). This wrapping  around $S_1$ parametrized by $x_4$
   leads to     vanishing action at the tip of the cigar geometry at low temperature. The D2 branes responsible for the transitions 
    $  | n\ra \rightarrow  | n+1\ra$  precisely satisfy both these properties, which explains our choice for a relevant configuration in the dual description.   
     As we mentioned at the end of section \ref{model} one type of the 
   D2 branes was identified earlier in ref. \cite{Gorsky:2007bi}  as magnetic strings. Another type of the D2 branes was identified as D2 domain walls~\cite{Yee}. The D2 magnetic strings carry 4d topological charge by construction, while D2 domain walls carry $Z_N$ charge~\cite{Yee}. Nevertheless, these D2 domain walls in combination with D0 branes  in strong coupling regime at $T<T_c$ are capable to carry 4d topological charge. Indeed,  the corresponding D0-D2 configurations can be explicitly constructed   in some supersymmetric models \cite{Davies:1999uw} where analytical QFT based computations can be also  performed. Similar  D0-D2 configurations may also emerge in strongly coupled QCD at $T<T_c$ as argued  in \cite{Gorsky:2009me}.
  
   Therefore, in what follows we shall not discriminate   between these two options in our discussions as lattice studies can not presently 
    precisely determine the dimensionality of fluctuating quantum objects. In particular, 
in ref. \cite{Buividovich:2011cv} it has been argued that these long range configurations actually might be characterized by Hausdorff dimension which gradually varies with  cooling procedure. 
  Therefore, we shall 
   use a single  term ``D2 brane"  to describe these objects in spite of the fact that these configurations are physically different. 
   We expect that some future, more refine lattice  studies will allow to discriminate between  those two options (or rule out both of them)\footnote{I am  thankful to Ivan Horvath for his comments on present measurements of dimensionality of the objects observed on the lattice simulations, and on future progress which should be expected as a result of these measurements.}.

  The recent 
Monte Carlo studies of Yang Mills  theory  have revealed a laminar structure in the vacuum consisting of extended, thin, coherent, locally low-dimensional sheets of topological charge embedded in 4d space, with opposite sign sheets interleaved, see original QCD lattice results~\cite{Horvath:2003yj,Horvath:2005rv,Horvath:2005cv,Alexandru:2005bn}. 
One should add that a similar structure has been also 
observed in QCD by a different groups~\cite{Ilgenfritz:2007xu,Ilgenfritz:2008ia,Bruckmann:2011ve,Kovalenko:2004xm,Buividovich:2011cv}. It has been also observed    in two dimensional $CP^{N-1}$ model~\cite{Ahmad:2005dr}.   
We do not make any attempts in this work 
  to cover this   subject which has  a number of subtle points. Instead, we 
  concentrate 
 on  very few key properties of the gauge configurations which apparently make crucial contributions to the $\theta$ dependent portion of the energy expressed in terms of the  topological susceptibility (\ref{YM}). One can argue that such a complex structure observed on the lattice  can be  thought as a result of dynamics of the D2 branes  described above\footnote{Similar  idea with interpretation of the D6 branes  in   holographic description being responsible for topological susceptibility     was recently  advocated in ref.\cite{Thacker:2011sz}.  One should remark that these D6 domain walls  are in fact $\theta$-independent  configurations according to the   classification presented in section \ref{model}, and they can not interpolate between different  topological sectors  $| n\ra$ and  $| n+1\ra$. 
    In different words, these D6 domain walls  interpolate between the ground state and an excited  local minimum characterized  by  a small extra energy density $\Delta E\sim 1/N_c$. 
  These domain walls  are  similar to the physical static domain walls in ferromagnetic systems interpolating between physically distinct vacua.
  These D6 domain walls have finite action and finite width in contrast with thin tensionless D2 branes.     Such domain walls  in QCD within QFT approach in the presence of light quarks  in large $N_c$ limit when   the $\eta'$ is light
$m_{\eta'}^2\sim 1/N_c$
  had been  previously studied  in~\cite{Forbes:2000et}.}.

 Indeed, it has been observed that  
the tension of the ``low dimensional objects"  vanishes below the critical temperature and these objects percolate through the vacuum, forming a kind of a vacuum condensate. 
Furthermore, 
these ``objects"  do not percolate through the whole 4d volume, but rather, lie on low dimensional surfaces $1\leq d < 4$~\cite{Horvath:2005rv}.  The total area of these surfaces is dominated by a single percolating cluster of ``low dimensional object". From holographic perspective   these Monte Carlo  results fits very nicely with conjecture that the observed   structure 
can be identified with   the D2 branes described above. In particular the tension of the D2 branes vanishes 
in confined phase  as a result of cigar geometry in the holographic 5d description such that the D2 brane shrinks to the tip as we already mentioned.
Therefore, an arbitrary large number of D2 branes  could be produced, resulting in their condensation. 
 The observed percolation (condensation) of these objects in lattice simulations unambiguously  imply that they must have vanishing effective tension.
Otherwise, only finite, not infinite percolating  clusters could be observed, in contrast with observations. 

Another measurement  suggests that  the contribution of the percolating objects  to the topological susceptibility $\chi_{YM}$ has the same sign     compared to its total value. Furthermore,  both contributions have  opposite sign in comparison with any propagating physical states (\ref{G}), (\ref{top2}). 
These lattice studies  are consistent  
with    non-dispersive nature  of  D2 branes, which can not be related to any propagating physical degrees of freedom. It unambiguously implies that the   D2 branes should be thought as 
configurations describing the tunnelling events in Minkowski space, rather than as a real physical domain walls.

Furthermore, the width of the ``objects"  apparently vanishes in the continuum limit.
   These measurements, again,   support the 
tunnelling interpretation of the contact term. Indeed, it is well known property of conventional quantum mechanics that 
the   individual photons penetrate an optical tunnel barrier with an effective group velocity considerably greater than the vacuum speed of light~\cite{Steinberg:1993zz}, see also review \cite{Chiao:1998ah}. In different words, the  
tunnelling  time can not be distinguished  (experimentally)  from zero. In quantum field theory context it means that 
the configurations responsible for tunnelling events must have vanishing size (in continuum), which is precisely what has been measured
on the lattices.  The same feature  can be also seen 
on   Fig 5 of ref.\cite{Bruckmann:2011ve} 
 where the contact term with vanishing size (in continuum) has a ``wrong sign" in Euclidean lattice simulations. This again supports the tunnelling interpretation of this term in Minkowski space.

\subsection{ Variation of the contact term with small variation of the background}\label{corrections}
Our previous section \ref{lattice} was devoted to the discussions of the contact term in QCD from holographic perspective. 
Essentially, we have  not produced any new results in that section as the presence of the contact term is well established phenomena (supported by the lattice studies). We simply suggested a new interpretation of this term. However, the new interpretation may lead to some new physical results as we shall argue below. 

Main question we address here is related to a possible variation of the contact term when the background is slightly changed.
For example, we wish to know what happens when infinite time coordinate $\tau\in (-\infty, +\infty)$  is replaced by a finite size ring $\tau\in (0, 2\pi R$), or what happens when the Minkowski space-time $R_{3,1}$   is replaced by FRW metric characterized by dimensional parameter $R\sim H^{-1}$ describing the size of the visible universe  with $H$ being   the Hubble constant. 

Normally, one should not expect any strong dependence on very large distances 
 as QCD has a mass gap $\sim \Lqcd$. Therefore, one should expect the exponential suppression $\sim \exp(-\Lqcd R)$ for any local observables. Formally, this expectation follows from the dispersion relation similar to (\ref{G}) written in coordinate space which explicitly shows an exponentially weak sensitivity $\sim \exp(-\Lqcd R)$  to large distances. 
 The main point of this paper is  that along with conventional dispersive contribution (\ref{G}) there is also non-dispersive contribution (\ref{top1}) which is not related to any physical propagating degrees of freedom, and which can not be computed using the dispersion relations. 
 This contribution as we discussed above  emerges  as a result  of topologically nontrivial sectors in four dimensional QCD, and it  may lead to a power like 
corrections $ R^{-p}$ with power $p\geq 1  $ rather than exponential like $  \exp(-\Lqcd R)$.   In fact,  this term 
in our framework  is described by topologically protected massless ghost field (\ref{top3}) as discussed in section \ref{lattice}. Therefore, in this framework, it is indeed quite natural to expect  a power like  corrections $ R^{-p} $.

We want to get some insights on this problem from the holographic perspective. We start    our discussions   with few general remarks. 
In principle, when we slightly change the geometry, one should 
compute the corresponding changes in the vacuum energy as a result of changes of the boundary conditions. The corresponding changes  represent  the Casimir vacuum energy in multidimensional space   due to new specific boundary conditions which were trivial ones in previous computations (\ref{du}), (\ref{laplace})   with   infinite Minkowski space-time background.  
 Technically, it could be very challenging problem as it requires the computation of a corresponding multidimensional Green's function in curved background which satisfies specific boundary conditions. However, on a general ground we expect that the contact term (\ref{du}) receives a  power-like correction $\sim \epsilon^p$, not exponentially small corrections $\sim \exp(-1/\epsilon)$
if deformation of the geometry parametrized by small parameter $ \epsilon\sim (R\Lqcd)^{-1} \ll 1$. 

Indeed, the computations of the Casimir vacuum energy in 5-dimensional  space $(x_{\mu}, u)$ can be  accomplished by using effective Lagrangian describing  the {\it massless}  RR (\ref{RR})   or axion field\footnote{We use term ``axion" which is 
a jargon here.  There is no real new dynamical degree of freedom in 4d space such as axion, see  reviews~\cite{axion} about  the  physical 4d axion 
which is  as real degree of freedom and the dark matter  candidate. However,  the $\theta$ dependent portion of the vacuum energy in holographic picture 
is determined  
by the dynamics of the axion field living in multidimensional space. Precisely the dynamics of the axion field determines 
the magnitude of the integral (\ref{du}) which corresponds to the physically observable topological susceptibility $\chi_{YM}$.
This relation  of the axion field  from multidimensional space  with $\chi_{YM}$ justifies  our terminology.}
living in the bulk of multidimensional space. This field $C_1$   is responsible for the contact term (\ref{du}). More importantly, the axion field is massless
in multidimensional space as a consequence of corresponding gauge invariance. Therefore, while a computation of the Casimir vacuum energy in 5-dimensional  space $(x_{\mu}, u)$  could be very challenging technical problem, an estimation of  this energy  is simple: we expect a power -like correction to the vacuum energy as a   natural consequence of the dynamics of  a {\it massless}  axion   field
living in the bulk of multidimensional space,  which is responsible for the contact term.

We will   test this argument (on power like correction) below 
by replacing infinite Euclidean time coordinate $\tau\in (-\infty, +\infty)$  by a finite size ring $\tau\in (0, 2\pi R$). As is known, sufficiently small parameter $R$ corresponds to    large temperatures $T\gg T_c$ where QFT based computations 
are justified. After that we apply the same argument in strong coupling regime where QFT based analytical computations are not available.
Nevertheless, we will quote  some recent lattice numerical results supporting our argument.

 First, we consider  de-confined phase when the  length of the ring $S^1_{\tau}$ is quite small and order of $ T_c^{-1}$.
The contact term is expected to be exponentially suppressed at $T>T_c$  in de-confined regime as D0 brane wrapping around $x_4$  has a finite tension,   saturates the topological susceptibility, and leads to the exponentially small contact term~\cite{Bergman:2006xn,Parnachev:2008fy}.

This picture can also  be easily  understood from  QFT viewpoint~
\cite{Parnachev:2008fy,Gorsky:2009me}. Indeed,
the wrapping around $x_4$  corresponds to the well
 defined small instanton and one can use the standard instanton calculus to estimate
 the critical temperature $T_c$ and the $\theta$ behaviour above $T_c$:
  \be
   \label{T}
   V_{\rm inst}(\theta, T>T_c) &\sim& \cos\theta \cdot e^{-\alpha N_c\left(\frac{T-T_c}{T_c}\right)}, ~~~~ 1\gg \left(\frac{T-T_c}{T_c}\right)\gg 1/N_c,
    \nonumber \\
    \chi_{YM} (T>T_c) &\sim& \frac{\partial^2 V_{\rm inst}(\theta, T)}{\partial\theta^2}\sim
      e^{-\alpha N_c\left(\frac{T-T_c}{T_c}\right)}\rightarrow 0,  ~~~ \alpha \sim 1, ~~~ N_c\gg 1.
   \ee
  Such a behaviour    implies that the dilute instanton gas approximation at large $N_c$ is
justified even in close vicinity of $T_c$ as long as $\frac{T-T_c}{T_c}\gg \frac{1}{N_c}$.
  Such a sharp drop  of the topological susceptibility $ \chi_{YM} (T)$
is supported by  numerical lattice results \cite{Vicari:2008jw} which unambiguously
suggest that the topological fluctuations are strongly suppressed in  de-confined phase,
and this suppression becomes more severe with increasing $N_c$.  The behaviour (\ref{T}) in de-confined phase should be contrasted with
formula (\ref{du}) corresponding to the  confined phase when  $ \chi_{YM} (T=0)\sim N_c^0$.

Now we want to  address the following question: how  does the contact term $\chi_{YM}$ vary  when the  radius of the ring $S^1_{\tau}$ slightly changes?
We want to study these changes by remaining  in a deep de-confined phase $(T\gg T_c) $, such that no drastic variations  are expected to occur.
One can easily answer this question by noticing that small variation in $\tau\rightarrow \tau+\Delta \tau$ corresponds to a small variation of the temperature $T\rightarrow T-\Delta T$ with $\Delta T= (2\pi \Delta \tau)^{-1}$ while keeping $T_c$ unchanged. Therefore, a small variation of the
background $\tau\rightarrow \tau+\Delta \tau$ results in a small (power-like) variation of the contact term
\be
   \label{delta}
    \chi_{YM} (T-\Delta T)  \sim 
      e^{-\alpha N_c\left(\frac{T-\Delta T-T_c}{T_c}\right)} \sim    \chi_{YM} (T)\cdot \left[ 1+\alpha N_c\cdot \frac{\Delta T}{T_c}\right], ~~~~~~ \Delta T/T_c\ll 1, ~~  T_c \ll  T.
   \ee
Therefore, a small increase of the the  radius of the ring $S^1_{\tau}$ leads to a small increase of the contact term reflected by eq. (\ref{delta}). 
 The contact term  obviously receives a power like correction $\sim\epsilon$ when the geometry is   slightly  changed $\sim\epsilon$, 
\be
   \label{correction}
    \frac{\left[\chi_{YM} (T-\Delta T)  -   \chi_{YM} (T)\right]}{ \chi_{YM} (T)} \sim \epsilon, ~~~{\rm where} ~~~
     \epsilon\equiv \frac{ \Delta  \tau}{\tau}\sim \frac{\Delta T}{T}, ~~~~  \Delta  \tau\ll \tau .
   \ee
   The  correction (\ref{correction}) obviously supports our general argument presented above on power like corrections $\sim \epsilon^p$
   when manifold slightly changes as the variation of the manifold is explicitly translated in changes  of the temperature in conventional Euclidean formulation. 
   
   Now, we want to study a similar variations of the contact term  $\chi_{YM} (T-\Delta T)$  with slight change of  radius of the ring $S^1_{\tau}$ but in confined phase  with $T\ll T_c$. This corresponds to a strong coupling computations when no  analytical formula similar to (\ref{T})  exists in this regime. 
 From the holographic prospective, however,  the the result must be very similar to the previously discussed case (\ref{correction}),  as the  contact term (\ref{du}) and corresponding vacuum energy,   receives the power like corrections  $\sim \sum_{p\geq 1}c_p\epsilon^p$, not exponentially small corrections $\sim \exp(-1/\epsilon)$ as all  changes  are governed by the 
{\it massless}  axion  field  field living in the bulk of multidimensional space.     This argument is consistent with  the direct lattice computation
 \cite{Holdom:2010ak}   which indeed finds a power like corrections $ R^{-p}$ with $p\simeq 2.34 $ for smallest lattice size $R=12$ and    $p\simeq 1.33$ for the largest available lattice size $R=24$ in lattice units. Such a behaviour should be contrasted with naive expectation
  $  \exp(-\Lqcd R)$ based on conventional dispersion relations without an accounting for the non-dispersive term. 
  
  We should emphasize that our general  argument does not specify the power   $p$  for the Casimir-like correction $\sim \epsilon^p$. 
In general, one  could expect any integer number $p\geq 1$ as   QFT based computations would  suggest. Indeed, if 
  2d QED model is defined on a generic torus determined by Teichm\"uller complex parameter $\tau\equiv \tau_1+i\tau_0$ the correction  behaves linearly in inverse size,  $\sim R^{-1}$.  The suppression  becomes  much stronger $\sim  R^{-2}$ (but still, not exponential)  in special symmetric case when metric $g_{\mu\nu}\sim {\rm diag} (1, 1) $  ~\cite{Urban:2009wb}.  Also, computations of the vacuum energy in 2d $CP^{N-1}$ model  defined on a finite interval of length $R$ with Dirishlet boundary conditions at large $N$ also exhibits  the correction    $\sim R^{-1}$ \cite{Milekhin:2012ca}.  
  A similar effect occurs in ``deformed QCD" model discussed in section \ref{deformed} when the theory is formulated on 
$S^1\times   R^3$ and  infinitely large space $R^3$  is replaced by a sphere $ S^3$ with  radius R. The correction behaves in this case as $R^{-1}$ \cite{Thomas:2012ib}.  One should expect $R^{-2}$ correction for a  symmetric  infrared regularization when  infinitely large space $R^4$  is replaced by a sphere $ S^4$ 
 \cite{Thomas:2012ib}.  We suspect  that a similar variation of power $p$  would also emerge in holographic description, however the corresponding study is beyond the scope of the present work. The experience with QFT computations suggest that a generic asymmetric background leads to a minimal $p=1$ suppression, while more symmetric cases lead to a higher power of suppression $p\geq 2$.  To conclude:  all these simple calculable models with a gap exhibit the power like corrections, in contrast with  naive expectation that there should   exponentially weak 
 $ \sim  \exp(- R)$ sensitivity to a size of the system $R$.

  \vspace{0.1cm}
{\it\underline{Few more comments:}}
\vspace{0.1cm}

We want to avoid any confusion with  the terminology. In what follows, 
we shall use the term ``topological Casimir effect" to discriminate it from conventional Casimir effect when power like behaviour is due to the real massless  propagating degrees of freedom living in 4 dimensional space, such as $E\&M$ photons, in huge contrast with our case when there are no   any massless asymptotic states in QCD. 
As we mentioned previously, this ``topological Casimir effect" occurs in gauge theories when nontrivial ``degenerate"  topological sectors are present in  the system.   For the present work it is important that  the same ``topological Casimir effect" in 4d can be thought from  the dual holographic description
as conventional Casimir effect in 5 dimensional space as a result of  massless    axion field living in the bulk of multidimensional space.

 The ``topological Casimir effect"  obviously  is very unnatural and unexpected effect which   begs for a simple intuitive explanation. A formal explanation was given in  the text and was formulated in terms of the topologically protected pole 
 at $k^2=0$, see eq. (\ref{top3}) within QFT framework. A similar formal explanation  in dual holographic  description is given in terms of the  massless  axion field  living in the bulk of multidimensional space.  
As this  effect plays a crucial role in the application considered in section \ref{consequences},  we want to present  here few other systems    where a similar phenomena occurs, and  where it has precisely the same nature. 
 Furthermore, in these systems a similar  problem  can be exactly solved (in drastic contrast with  strongly coupled 4d QCD). Most importantly,   analogous   effects  in these systems have been   experimentally observed. 
  
  Our first example is from quantum mechanics (QM) and it is the well known  Aharonov -Casher effect as formulated in \cite{Reznik:1989tm}.
The relevant part of this work can be stated as follows. If one inserts an external charge into superconductor when the electric field is exponentially suppressed $\sim \exp(-r/\lambda)$ with $\lambda $ being the penetration depth,  a neutral magnetic fluxon will be still sensitive to an inserted external charge at arbitrary large distance. The effect is pure topological and non-local in nature.  The crucial element why this phenomenon occurs in spite of the fact that the system is gapped is very similar to what we discussed in the present work. First of all, it is the presence of different topological states  $u_n$ (number of Cooper pairs) in the system and ``tunnelling" between them  (non-vanishing matrix elements  between $u_n$ and $u_{n+1}$ states) as described in  \cite{Reznik:1989tm}. Those states are  analogous  to the topological sectors $|n\ra$ in our work.
As a result of the ``tunnelling", an appropriate ground state $U(\theta)$ must be constructed as discussed  in \cite{Reznik:1989tm}, analogous to the  $|\theta\ra$ vacuum construction in gauge theories.
This state $U(\theta)$ is an eigenstate of the so-called ``modular operator" which commutes with the hamiltonian. In our work an analogous role   plays the large gauge transformation operator $\cal{T}$ such that $ {\cal{T}}|\theta\ra=\exp(-i \theta )| \theta\ra$. An explicit construction of the operator  $\cal{T}$ is known: it is non-local operator   similar to non-local ``modular operator" from ref. \cite{Reznik:1989tm}, see Appendix \ref{tau} for some technical details. The crucial element of the construction of ref.\cite{Reznik:1989tm} is that the induced charges in presence of the gap can not screen the ``modular charge" as a result of commuting the ``modular operator" with hamiltonian.  This eventually leads to a non-vanishing effect, i.e.  collecting a non-zero phase at arbitrary large distance. 
 We do not have a luxury to resolve the problem using the hamiltonian description in  strongly coupled four dimensional QCD. However,  one can argue that  the role of ``modular operator"
is played by large gauge transformation operator $\cal{T}$ which also commutes with the hamiltonian $[{\cal T},H]=0$, such that our system must be  transparent to   topologically nontrivial pure gauge configurations, similar to transparency of the superconductor to the ``modular electric field" from ref. \cite{Reznik:1989tm}. Such a behaviour of our  system can be thought as a non-local topological effect
  similar to the   non-local Aharonov -Casher effect as formulated in \cite{Reznik:1989tm}, see also few comments in Appendix \ref{tau} on this similarity. The last word whether this analogy can be extended to the strongly coupled four dimensional QCD remains, of course,   the prerogative of the direct lattice computations similar to recent studies \cite{Holdom:2010ak}.

 Our second example is description of a superconductor using the so-called topological ``BF" action as presented in \cite{BF}.
 This QFT description  deals exactly with the same question on  how does a moving vortex detect a stationary charge, given that the electric field is exponentially screened. Paper  \cite{BF} explicitly shows how pure gauge (but topologically nontrivial) field 
 may lead to long range effects. In many respects this description is very similar to our description in terms of unphysical  ghost degrees of freedom when the topological features of the system are represented by auxiliary fields which are responsible for the dynamics of the degenerate ground state. 
 Our final example is description of topological insulators in terms of topological ``BF" action as presented in \cite{Cho:2010rk}.
 In this case, again, some pure gauge, but topologically nontrivial auxiliary fields may penetrate into  gapped insulator to produce  an effective massless mode on the surface of a sample. Such a behaviour of the system can be thought as a non-local topological effect similar to the  previously considered example~\cite{Reznik:1989tm}.
 
 We conclude this section with the following comment. Our interpretation of  dynamics of the system, when it is  described in terms of the different  topological sectors $|n\ra$
 and large gauge transformation operator $\cal{T}$  commuting with the hamiltonian $[{\cal T},H]=0$,   is obviously a gauge variant interpretation, see Appendix \ref{tau} with some technical details. As we already mentioned previously, there is no any physical degeneracy in QCD, in contrast with condensed matter examples   when distinct physical degenerate quantum states are present in the system. 
 Nevertheless, such gauge-dependent  interpretation  helps us to understand a number of very nontrivial features of the contact term which are known to be present 
 in lattice Monte Carlo simulations.  
 
 If, instead, we consider a physical Coulomb gauge when all unphysical degrees of freedom are removed from the system, the corresponding physics related to nontrivial topological structure of the gauge fields does not go away.  Instead, it will reappear in terms of the  so-called Gribov copies leading to a strong infrared (IR) singularity.  Precisely this IR behaviour due to the Gribov  copies was the crucial element in the approach advocated in ref.\cite{Holdom:2010ak}  where power -like correction  $R^{-p}$  has been observed in lattice simulations. 
 Similarly, the construction of the ground state  in exactly solvable 2d QED in physical Coulomb gauge,  as originally discussed in \cite{KS}, 
  is characterized by long range forces. This long range force  prevents distant regions from acting  independently. This is  essentially 
  the same  manifestation of long range forces  in  physical Coulomb gauge in the absence of  unphysical ghost. We opted  to interpret the results in  gauge-dependent framework in terms of   different  topological sectors $|n\ra$
 and large gauge transformation operator $\cal{T}$  where one can use analogy with condensed matter systems. Alternative  option is lattice numerical simulations where entire  notion of   vacuum winding states  $|n\ra$ and large gauge transformation operator $\cal{T}$  do not even exist
 as there is a unique ground state. However, irrespectively to the interpretation, the last word in strongly coupled QCD is expected from the direct lattice simulations 
 which eventually should confirm or rule out our arguments suggesting that the sensitivity to arbitrary large distances should be power like $R^{-p}$ rather than  exponential  like  $\exp(-\Lqcd R)$ as a consequence of nontrivial topological properties of strongly coupled QCD. 
 
  The most important result of this section which  plays a crucial role in the  application considered in next section can be formulated as follows. 
 The contact term (and corresponding $\theta$ dependent portion of the vacuum energy density)  may receive  power like corrections $\sim R^{-p}$  in QCD
 despite the presence of a mass gap in the system.  From 4d viewpoint this is the 
   ``topological Casimir effect"  when no massless degrees of freedom are present in the system, and effect within QFT framework can be explained 
   as a result of degenerate topological sectors in the theory  represented by  topologically protected unphysical pole  
 at $k^2=0$, see eq. (\ref{top3}).  The same effect from 5d holographic  description can be explained in terms of the  massless  axion field  living in the bulk of multidimensional space.  From 5d viewpoint this is a ``conventional Casimir effect"  due to a physical (in multidimensional space) massless  axion.  
   
\section{Contact interaction in QCD and profound consequences for expanding  universe}\label{consequences}

This portion of the paper  which is the direct manifestation of the ``topological Casimir effect" discussed in previous section is much more speculative in nature than the previous sections. The corresponding speculation on profound consequences 
for cosmology  of this effect have been already formulated in author's previous works.
  However, there are two new elements on   application of 
  the ``topological Casimir effect" to cosmology
 which were not discussed previously, and will be  elaborated here.   First of all, however,  we need to   make a short overview  of this  proposal.
   
 The  $\theta-$ dependent portion of the  energy which is 
not related to any physical propagating degrees of freedom,  is well established effect. It has been   tested a numerous number of times in the lattice simulations. How does this energy change when  the background slightly varies?  The main motivation for  this question is as follows. 
We adopt   the paradigm that the relevant definition of  the energy  which   enters  the Einstein equations 
 is the difference $\Delta E\equiv (E -E_{\mathrm{Mink}})$, similar to the well known Casimir effect.    Such a definition of the ``physical energy"  in fact is   used in  description of the horizon's thermodynamics \cite{Hawking:1995fd,Belgiorno:1996yn} as well as  in a course of computations of different Green's function in a curved background~\cite{Birrell:1982ix}. 
  In the present context  such a definition $\Delta E\equiv (E -E_{\mathrm{Mink}})$ for the vacuum energy for the first time was advocated   in 1967   by Zeldovich~\cite{Zeldovich:1967gd},  see also 
  \cite{Zhitnitsky:2011tr} with  a large number of related  references.

As we mentioned previously,  a naive expectation   suggests that 
$\Delta E \sim \exp(-\Lqcd/H)\sim \exp(-10^{41})$ as QCD has a mass- gap $\sim \Lqcd$, and therefore, $\Delta E$ must not be sensitive to size of our universe.
In this estimate we  do not distinguish    the  size of the visible universe $\sim H^{-1}$  
from   size $R$ of a compact 
manifold we used in  our analysis   in previous section. A crucial distinct feature which is important for our discussions     is the presence of dimensional parameter $R\sim H^{-1}$   in a system which discriminates it  from infinitely large  Minkowski space-time.     

 This naive argument, however,  may fail as a result of nontrivial  topological properties of QCD  as we discussed above.   It   may lead to a power like 
scaling $\Delta E\sim H^p$ rather than exponential like $\Delta E \sim \exp(-\Lqcd/H)$.     
From   holographic viewpoint such power like scaling 
   $\Delta E\sim H^p$   follows from the fact that the contact term 
  (or, what is the same, the $\theta$ dependent portion of the energy) is determined by massless  axion  field. It is naturally to expect that massless axion field produces  power like corrections in holographic description as discussed  in section \ref{corrections} such that    surface term (\ref{du}) which determines the magnitude of the contact term  
 receives a power like corrections when the background is slight modified. From holographic perspective, one can view this power like correction as a conventional Casimir effect in multidimensional space. However, in our four dimensional space this correction should be interpreted  as  the ``topological Casimir effect" as there are no any physical massless asymptotic states    in gapped QCD.
  
The power like scaling $\Delta E\sim H^p$ with $p=1$ (as some explicit QFT  computations suggest, see section \ref{corrections})  may have some profound consequences for evolution of our universe. 
  If true, the difference between two metrics (expanding universe  and Minkowski space-time) would lead to an estimate 
   \be
   \label{Delta}
   \Delta E\sim H\Lqcd^3\sim (10^{-3} {\text eV})^4,
   \ee
which is amazingly close to the observed DE value today. Furthermore, the power like scaling  (\ref{Delta}) with $p=1$ would imply that our universe approaches the de-Sitter state with constant expansion rate $H_{\infty}\simeq G \Lambda_{QCD}^3$ in asymptotic future \cite{Zhitnitsky:2011tr}.  
Such a behaviour $\Delta E\sim H +{\cal O} (H)^2$   was   postulated in \cite{dyn}.   
 This proposal has received some theoretical QFT based support  in \cite{Zhitnitsky:2010ji,Zhitnitsky:2011tr,Thomas:2012ib} where it was argued that power like  scaling indeed may emerge  as a  
  result of non-dispersive nature  of contact term (\ref{top1}), in contrast with conventional dispersive term (\ref{G}) related to
  physical propagating degrees of freedom. This correction does not violate unitarity,  causality and other important QFT properties. 
  
  We arrive to the same conclusion  in the present work   using the dual holographic arguments. 
  In the dual picture the DE in this framework is due to the Casimir effect in multidimensional space as a result of the  dynamics of the massless axion field in expanding universe. From  four dimensional view point this effect should be interpreted as a ``topological Casimir effect" when no massless asymptotic states are present in the system.  
  \exclude{
   It is interesting to note that 
expression (\ref{Delta})  reduces to 
Zeldovich's formula   $\rho_{\text{vac}} \sim Gm_p^6 $ 
   if one replaces $ \Lambda_{QCD} \rightarrow m_p $   and $  H\rightarrow G \Lambda_{QCD}^3$. 
   The last step follows from 
      the  solution of the Friedman equation 
     \be
 \label{friedman}
 H^2=\frac{8\pi G}{3}\left(\rho_{DE}+\rho_M\right),  ~~ \rho_{DE}\sim H\Lqcd^3
  \ee    
  when the DE component dominates the matter component, $ \rho_{DE}\gg\rho_M$. In this case    the evolution of the universe  approaches a  de-Sitter state with constant expansion rate $H\sim G \Lambda_{QCD}^3$ as follows from (\ref{friedman}).
  }
   A comprehensive phenomenological analysis of this model has been recently performed in ~\cite{Cai:2010uf} where comparison 
 with current observational data including SnIa, BAO, CMB, BBN has been presented, see also \cite{ohta,Sheykhi:2011xz,RozasFernandez:2011je,Cai:2012fq, Saaidi:2012av,Feng:2012wx} with related discussions.  The conclusion  was that the  model (\ref{Delta}) is consistent with all presently available data, and we refer the reader to these papers on analysis of the observational data.

 \section*{Conclusions}

 The main result of this work can be formulated as follows. We studied a number of different ingredients related to $\theta$ dependence, the non-dispersive contribution in topological susceptibility with the ``wrong sign", topological sectors in gauge theories, and many related subjects.
We argued that the corresponding physics in holographic description should be interpreted as a result of tunnelling events 
governed by the dynamics of the D2 branes. We quoted  a number of theoretical as well as numerical results supporting this interpretation. Essentially, we interpreted the holographic formula (\ref{du}) in terms of a singular behaviour of the contact term (\ref{YM}). The diverging nature of the core and vanishing width (in the continuum limit)  of this non-dispersive contribution to $\chi$  is supported by recent lattice studies.  After all, it is not really a huge surprise that a vanishing width of the contact term (confirmed by the lattice studies) is saturated by the objects (also observed on the lattices) which  have vanishing sizes  and which lead to a superluminal  behaviour (when interpreted  in Minkowski space-time), see original lattice results in \cite{Horvath:2003yj,Horvath:2005rv,Horvath:2005cv,Alexandru:2005bn,Ilgenfritz:2007xu, Ilgenfritz:2008ia,Bruckmann:2011ve,Kovalenko:2004xm,Buividovich:2011cv}. 

However, we believe that the most important result of our work is not just a  new interpretation of  old  results \cite{witten,ven,vendiv} related to the resolution of the $U(1)_A$ problem in QCD. 
Rather, the holographic description allows us to  test the sensitivity of the gauge theory with non-trivial topological features to arbitrary large distances.   A naive expectation based on dispersion relations dictates that a sensitivity to very large distances must be exponentially suppressed when the mass gap is present in the system. However, we argued that along with conventional dispersive contribution there exists a non-dispersive contribution, not related to any physical propagating degrees of freedom. This non-dispersive (contact) term  with the ``wrong sign"  emerges in QFT -based framework  as a presence of  topologically nontrivial sectors and tunnelling events between them. Technically, it is formulated in terms of unphysical  massless ghost field  (\ref{top3}) effectively describing the dynamics of these tunnelling transitions. In dual holographic description the same  
  dynamics is described by   the massless axion field living in the bulk of multidimensional space. From 4d viewpoint, the variation of this contact term with variation of the background leads to a power like ``topological Casimir effect". The same effect, but viewed from 5 dimensional perspective can be thought as conventional Casimir effect      which is a consequence of dynamics of   the  axion massless field.  
  
   From 4d viewpoint the transparency of the  QCD vacuum  to topologically nontrivial  gauge configurations is  similar to transparency of a  superconductor to the ``modular electric field" in  the Aharonov -Casher effect as discussed at the end of section \ref{corrections}, see Apendix \ref{tau} with some technical details on this similarity. In both cases
  an exponential suppression of the effects is avoided as a result of topological properties of gauge theories, and can be interpreted as a nonlocal effect. 
  We suspect that there should be a holographic description of the Aharonov -Casher effect when its ``modular electric field" is represented by the axion -like massless field living in the bulk of multidimensional space. We leave this subject for the future studies.

The ``topological Casimir effect" in QCD, if confirmed by future analytical and numerical studies, may have profound consequences for understanding of the expanding  universe we live in.
Finally, what is perhaps more remarkable is the fact that some elements of this framework can be, in principle, experimentally tested in heavy ion collisions, including the Casimir like $L^{-p}$ scaling behaviour.  The parameter $L$ in heavy ion collisions describes a size of the region where the QCD vacuum is disturbed  as a result of collision. Dependence on $L$, in principle,  can be experimentally studied as   $L$ is related to  the  physical size of the  colliding ions,   see   \cite{Zhitnitsky:2010zx, Zhitnitsky:2012im} for the details.

\acknowledgments
 Author thanks the Galileo Galilei Institute for Theoretical Physics for the hospitality and 
 organization of the workshop ``Large N gauge Theories'' where this  project  was initiated.
  I   thank Sasha Gorsky  for the  discussions and collaboration  during  the initial stage of this project.
  I am also thankful to Ivan Horvath and Folk Bruckmann for correspondence on lattice results and comments  on feasibility to measure the Casimir like correction in Monte Carlo simulations.  I am also    thankful to  Dima Kharzeev, Larry McLerran,   and other members of Nuclear Physics groups at BNL and  Stony Brook U.     for useful and stimulating discussions related to the subject of the present work.
 This work was supported, in part, by the Natural Sciences and Engineering
Research Council of Canada. 

\appendix
\section { The large gauge transformation operator $\cal{T}$.}\label{tau}
The goal of this Appendix is to give a short overview  of properties of the  $\cal{T}$ operator which was essential element  in our discussions  of the physical interpretation  of the contact term. We also want to present some additional arguments demonstrating striking similarity between  $\cal{T}$ operator and   ``modular operator" in  the Aharonov -Casher effect as mentioned  at the end of section \ref{corrections}.

The basic  properties of the  $\cal{T}$ operator  were analyzed  in late seventies and we follow classical paper \cite{COO-2220-115} in description of some  features of this operator in the context of the present work. The starting point is the construction of the   unitary operator ${\cal{T}}$, 
\beq
\label{T1}
{\cal{T}}= e^{iQ_{\lambda}}, ~~~ Q_{\lambda}\equiv \int d^3x \dot{A}_i^a  \left(\nabla_i \lambda^a  +f_{abc}A_i^b\lambda^a \right), 
\eeq
where we literally use notations of ref.  \cite{COO-2220-115}. An arbitrary function $\lambda^a(x) $ in this formula is a c-function which is a parameter of the gauge transformations. In particular, if $ \lambda^a(x) $ vanishes  at spatial infinity one can represent the operator $Q_{\lambda}$ 
in form 
\beq
\label{T2}
  Q_{\lambda}=- \int d^3x \lambda^a (x) \left(\nabla_i \dot{A}_i^a  +f_{abc}A_i^b\dot{A}_i^c \right)=-  \int d^3x \lambda^a (x) C^a(A), 
\eeq
which obviously annihilates all physical states as the term in parentheses is just Gauss's law $C(A)|{\cal{H}}_{phys}\ra=0$. However, if  $\lambda^a (x) $ describes the large gauge transformation, the operator ${\cal{T}}$ becomes  nontrivial operator and describes the  transition between topologically different sectors of the theory:
\beq
\label{T3}
{\cal{T}} |n\ra =|n+1 \ra.
\eeq
In this case one should construct the so-called $|\theta\ra$ vacuum state which is an eigenstate of the ${\cal{T}}$ operator: 
\beq
\label{T4}
|\theta\ra =\sum_n e^{in\theta} |n\ra  , ~~~~~~~~~~~~{\cal{T}} |\theta\ra=  e^{-i\theta}  |\theta\ra .
\eeq
It is important  to emphasize that while operator ${\cal{T}}$ formally constructed as an operator of gauge transformations, this operator does change the state (\ref{T3}) as a result of global effect. Therefore, one should treat ${\cal{T}}$ as ``improper" gauge transformation 
(the ``large gauge transformation"). Still, ${\cal{T}}$ commutes with the hamiltonian $[{\cal{T}}, H]=0$. As we   mentioned in the text, there is no any physical degeneracy in QCD, in contrast with condensed matter examples   when distinct physical degenerate quantum states are present in the system.  
 Nevertheless, the  gauge-dependent  interpretation  in terms  of the tunnelling between $ |n\ra$ and  $|n+1 \ra$ states  is quite useful to develop  an intuitive picture of relevant physics, which otherwise hidden in the numerical lattice Monte Carlo simulations with  such puzzling elements as the ``wrong sign" contributions with  vanishing width as discussed in section \ref{holography}.

While this paper is mainly devoted to 4d QCD it might be instructive to present  the  large gauge transformation operator ${\cal{T}}_{2d}$  for 2d QED as well.  In this case, as it is known, the nontrivial topological structure emerges as a result of nontrivial mapping $\pi_1(U(1)) =\cal{Z}$ which replaces the classification of the topological sectors in QCD which was based on  $\pi_3(SU(N)) =\cal{Z}$ mapping.    In 2d  case a similarity between  the ``modular electric field" in  the Aharonov -Casher effect  and operator ${\cal{T}}_{2d}$ becomes much more apparent. Indeed, the 
large gauge transformation operator ${\cal{T}}_{2d}$  for 2d QED can be represented as follows
\beq
\label{T5}
{\cal{T}}_{2d}=   e^{-\frac{2\pi i}{g} E(\infty)} \cdot e^{\frac{i}{g}\int^{\infty}_{-\infty} dx\lambda(x) C(A) }
\eeq
where $g$ is the coupling constant in 2d QED, and $E=-\dot{A}_x$ is electric field. Formula  (\ref{T5}) was obtained  by integrating by parts the two-dimensional analog of  eq.  (\ref{T1})  and taking into account that the large gauge transformation in this case is characterized by $\lambda(x=\infty)-\lambda(x=-\infty)=2\pi$.
One can construct $|\theta\ra$ state similar to (\ref{T4}) with result that the expectation value of large gauge transformation operator ${\cal{T}}_{2d}$ is given by
\beq
\label{T6}
  \la \theta|{\cal{T}}_{2d}|\theta\ra=  e^{-i\theta}   , ~~~ \theta \equiv \frac{2\pi }{g} E_x(\infty).
\eeq
The fact that in 2d QED the electric field at infinity plays the role of the $\theta$ parameter of course is well known classical result~\cite{Print-76-0357 (HARVARD)}. 
A similarity of the expectation value (\ref{T6})  with  the Aharonov -Casher effect  as presented in ref.\cite{Reznik:1989tm} is striking as $\theta$ parameter in this case is almost identical  to the ``modular electric field" from ref.\cite{Reznik:1989tm}, with the ``only" difference that  $|n\ra $ states in 2d QED
are not physically distinct  degenerate states as we already emphasized, in huge contrast with physical degeneracy in condensed matter systems discussed at the end of section \ref{corrections}. Eventually, this difference leads to the ``wrong sign" in topological susceptibility as we argued in this work. The last word whether this analogy can be extended to the strongly coupled four dimensional QCD remains, of course,   the prerogative of the direct lattice computations similar to recent studies \cite{Holdom:2010ak}.

 \end{document}